# Co-Production of Cement and Carbon Nanotubes with a Carbon Negative Footprint

Stuart Licht
Chemistry, George Washington University, Washington, DC 20052, USA
slicht@gwu.edu

**Abstract**
C2CNT (Carbon dioxide to carbon nanotube) cement plants have been introduced and analyzed which provide a significant economic incentive to eliminate the massive $CO_2$ greenhouse gas emissions of current plants and serves as a template for carbon mitigation in other industrial manufacturing processes. Rather than regarding $CO_2$ as a costly pollutant, this is accomplished by treating $CO_2$ as a feedstock resource to generate valuable products (carbon nanotubes). The exhaust from partial and full oxy-fuel cement plant configurations are coupled to the inlet of a C2CNT chamber in which $CO_2$ is transformed by electrolysis in a molten carbonate electrolyte at a steel cathode and a nickel anode. In this high yield 4e- per $CO_2$ process, the $CO_2$ is transformed into carbon nanotubes at the cathode, and pure oxygen at the anode that is looped back in improving the cement line energy efficiency and rate of production. A partial oxy-fuel process looping the pure oxygen back in through the plant calcinator has been compared to a full oxy-fuel process in which it is looped back in through the plant kiln. The second provides the advantage of easier retrofit, while the first maximizes efficiency by minimizing the volume of gas throughput (eliminating $N_2$ from air).

    An upper limit to the electrical cost to drive C2CNT electrolysis is USD70 based on Texas wind power costs, but will be lower with fuel expenses when oxy-fuel plant energy improvements are taken account. The current value of a ton of carbon nanotubes is substantially in excess of a ton of cement. Hence a C2CNT cement plant consumes USD50 of electricity, emits no $CO_2$, and produces USD100 of cement and USD60,000 of carbon nanotubes per ton of $CO_2$ avoided. As the cement product ages, $CO_2$ is spontaneously absorbed. This is a powerful economic incentive, rather than economic cost, to mitigate climate change through a carbon negative process.

# Co-Production of Cement and Carbon Nanotubes with a Carbon Negative Footprint

Stuart Licht, Chemistry, George Washington University, Washington, DC 20052, USA, slicht@gwu.edu

Climate change is causing species extinction, harming the planet and humanity. Cement production accounts for the largest global emissions of carbon dioxide of any manufacturing process, generating 5-6% of the global anthropogenic emissions of this greenhouse gas. Humans consumes over 1 ton of concrete per person per year. $3 \times 10^{12}$ kg of cement annually, and the cement industry releases 9 kg of $CO_2$ for each 10 kg of cement produced; emitting 36 Gt globally annually [1,2]. In this study we propose and explore the conversion of cement plant carbon dioxide emissions to carbon nanotubes as an incentivized pathway to transform and eliminate $CO_2$ emissions from cement plants. Recently we've demonstrated a novel chemistry for the efficient, low cost, high yield transformation of $CO_2$ to carbon nanotubes [3-10] that builds on our $CO_2$ splitting studies [11-17]. In this process, voltage is applied to split $CO_2$ in an electrolysis chamber between a nickel anode and a galvanized steel cathode into pure oxygen gas and a solid carbon product. This is a low energy, high efficiency process when conducted in molten carbonate electrolytes, and provides a reaction pathway to transform the greenhouse gas into a high value commodity. $CO_2$ is bubbled into the carbonate, dissolves to form more carbonate as described in equation 1, and during electrolysis, oxygen is evolved at the anode, while a thick solid carbon builds at the cathode (Fig. 1). The carbonate electrolyte is not consumed and the net reaction is $CO_2$ splitting into carbon and $O_2$, for example using $Li_2CO_3$ as the electrolyte;

$$\begin{aligned}
\text{Dissolution:} \quad & CO_2(gas) + Li_2O(soluble) \rightarrow Li_2CO_3(molten) \\
\text{Electrolysis:} \quad & Li_2CO_3(molten) \rightarrow C(CNT) + Li_2O(soluble) + O_2(gas) \\
\text{Net:} \quad & CO_2(gas) \rightarrow C(CNT) + O_2(gas)
\end{aligned} \quad (1)$$

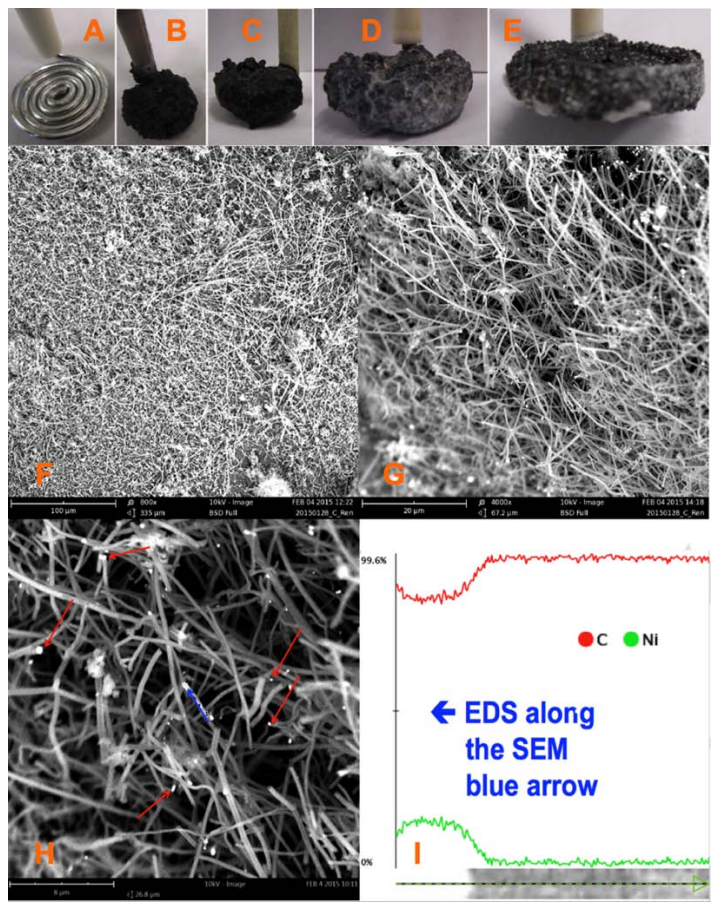

**Fig. 1**. $CO_2$ to carbon nanotubes formed at a coiled galvanized steel wire cathode with a nickel anode during 1 A constant current electrolysis in 750°C molten $Li_2CO_3$ electrolyte. SEM F to H are shown at various magnifications of the product removed from the cooled, washed cathode. "A" shows the 10 cm$^2$ coiled wire cathode prior to electrolysis. The anode is the inner wall of a 20 ml Ni crucible with electrolyte. "B-E" exemplify typical maximum variation of cathodes after a long (4Ah) electrolysis in molten carbonate. Red arrows in SEM "H" indicate typical Ni nucleation sites. The blue arrow originates at one Ni site and moves along the CNF path. "I": EDS composition mapping along the 6μ blue arrow path shown in SEM "H". Source: Licht, reference 3.

**The C2CNT process.** Under appropriate molten carbonate electrolysis conditions, and in accord with the net eq 1, $CO_2$ gas is converted to CNFs or their hollow form carbon nanotubes (CNTs). From their respective compositional mass, 1 ton of CNT is formed per 3.7 tons of $CO_2$ consumed. CNTs are stronger than steel, stable and compact, providing an ideal means to remove, transform and store $CO_2$ from flue gas. Carbonate's higher concentration of active, reducible tetravalent carbon sites logarithmically decreases the electrolysis potential and can facilitate charge transfer at low electrolysis potentials. We observe that carbonate is readily split to carbon approaching 100% coulombic efficiency (coulombic efficiency is determined by comparing the moles of applied charge to the moles of product formed, where each mole of solid carbon product formed depends on four moles of electrons) [3-10].

As shown in Fig. 2, $CO_2$ directly from the atmosphere or from $CO_2$ emissions is electrolyzed to produce a variety of valuable carbon nanomaterials that have a range of uses. The morphology of the carbon product will vary whether natural abundance $CO_2$ is used in the synthesis, which produces (hollow) CNTs, or whether $C^{13}$ is used, which produce (solid core) CNFs (Fig. 2B and 2C) [8]. High concentrations of lithium oxide can produce a tangled morphology (Fig. 2D), while no additional concentration of lithium oxide produces straight nanotubes (Fig. 2F). Fig. 2E shows cross-sections of the synthesized multiwalled carbon nanotubes, MWCNT. The use of naturally abundant $CO_2$, electrolysis current control, and the addition of small quantities of nickel to act as nucleating agents, leads to high yield of the particularly valuable CNTs. Due to their superior strength, conductivity, flexibility, and durability CNFs have a variety of applications including in nanoelectronics, in Li-ion batteries and as a principal component in the light-weighting of infrastructure construction materials, transportation (air, land, sea) vehicles, consumer electronics, wind turbine blades, and athletic equipment.

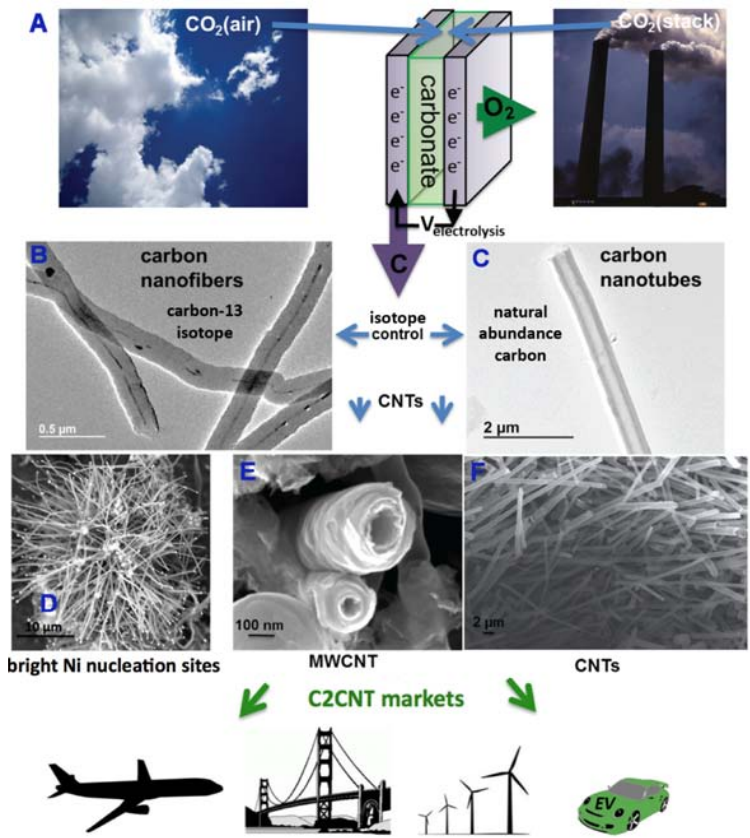

**Fig. 2.** Scheme for the electrolytic synthesis of carbon nanostructures from carbon dioxide: (a) the source of $CO_2$ can be either smoke stack concentrations of $CO_2$ or air (without requiring pre-concentration) dissolved in molten carbonate, (b) and (c) isotope controls formation of carbon nanotubes or nanofibers, the more valued CNTs are made from the less expensive natural abundance $CO_2$. High oxide concentration produce tangled morphologies (d) while low oxide produces straight nanotubes (f). (e) Bright nickel nucleation sites, as identified by EDS, initiate CNT growth. Lower panel: The CNTs and CNFs provide high conductivity and superior carbon composite lightweight structural materials for jets, bridges, wind turbines, and electric vehicle bodies and batteries. Source: Licht, reference 4 and 5.

High current densities (>1 amp cm$^{-2}$) of carbon formation are sustained, and we observe similar sustained currents at carbon, inexpensive steel cathodes, or expensive platinum electrodes (each cathode effectively become a carbon electrode during the product deposition). XRD, SEM, TEM and Raman spectroscopy confirm that the product is carbon nanotubes, and EDS elemental spectroscopy as shown in Fig. 1 confirms that the CNT ends contain transition metal (in this case nickel, nucleation points related to the CNT formation and growth). Fig. 1 presents a simple coiled galvanized steel wire (the cathode) prior to electrolysis, and following extraction after the electrolysis. The visible, voluptuous, thick black layer is CNTs mixed with congealed electrolyte. Simply uncoiling the wire allows the product to be retrieved for washing to remove the electrolyte, following which the wire can be recoiled and reused. The general process by which valuable carbon nanofibers and nanotubes are produced, rather than amorphous carbon or simple graphite, is summarized in Fig. 3.

Whereas molten $Li_2CO_3$ with trace concentrations of Ni, Cu, Fe or Co dissolve $CO_2$ to produce a high yield of straight CNTs (Fig. 1), the addition of oxide (such as $Li_2O$) yields tangled CNTs, which we've shown are useful for higher capacity rechargeable batteries. The tangling is due to a proliferation of sp$^3$ defects, rather than sp$^2$ (tetrahedral rather than honey-comb graphite structure), and these defects lead to not only twisting and tangling, but also enhanced Li-ion intercalation for improved rechargeable batteries. We've demonstrated that composite electrolytes such as Li-Na and L-Ba carbonates also produce a high yield of CNTs, and have advantages such as further reduced electrolyte cost or higher density electrolyte for ease of the CNT product extraction.

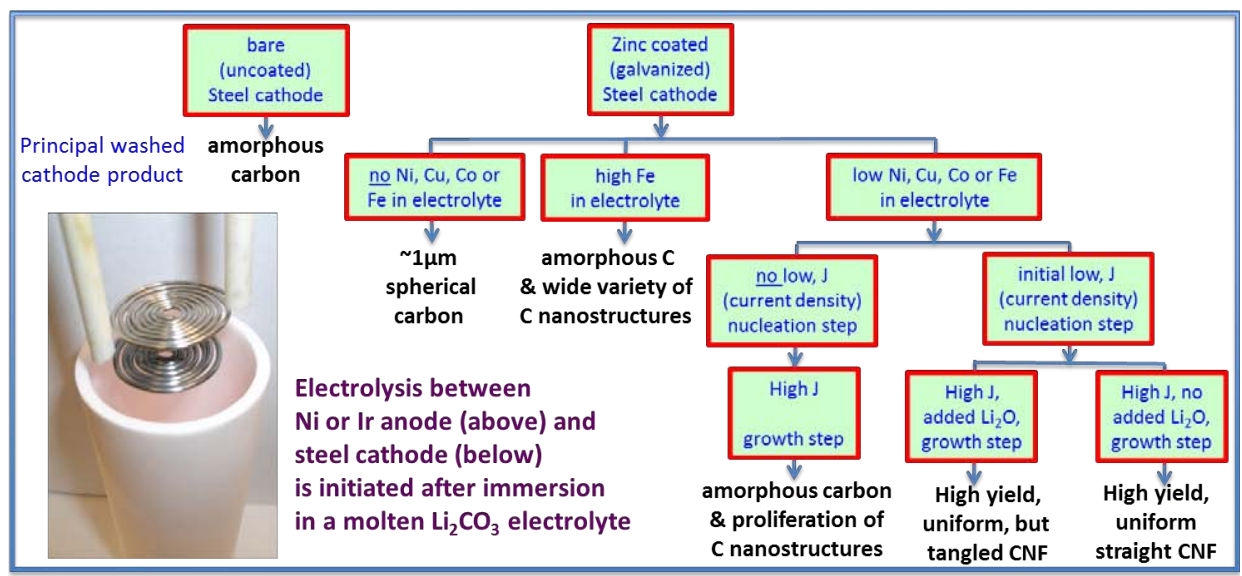

**Fig. 3**. Molten carbonate electrolysis pathways transforming $CO_2$ to high yield, CNF product. Reference 3.

Fig. 4 presents illustrates the C2CNT (carbon dioxide to carbon nanotube) process, and the increasing sized electrolysis chambers we are investigating, which to date have increased the electrode surface area capacity from 5 cm$^2$. The large cell shown provides an area of 0.8 m$^2$ per cell, and with a 0.2 A cm$^{-2}$ is capable of the generation of 4 kg of CNT per day while consuming 16 kg of $CO_2$ per day. In the electrolysis experiments, constant currents to ~100 A are provided by a Xantrex 8-100 A power supply and higher currents by a Power Ten Inc. power supply. In these step-wise scale-up experiments, 10 m$^2$ contiguous electrodes will utilize Ni clad on copper anodes and steel clad on copper anodes which evenly distribute current throughout the larger electrodes, to be capable of generating 50 kg of CNT per day and consuming 200 kg of $CO_2$ per day. Fig. 5 illustrates a two chamber C2CNT configuration in development which separates the electrolysis and $CO_2$ dissolution functions into two separate chambers and allows for ease of filtration removal of particulate impurities, and shows the related pumped molten salt electrolyte loop in the laboratory.

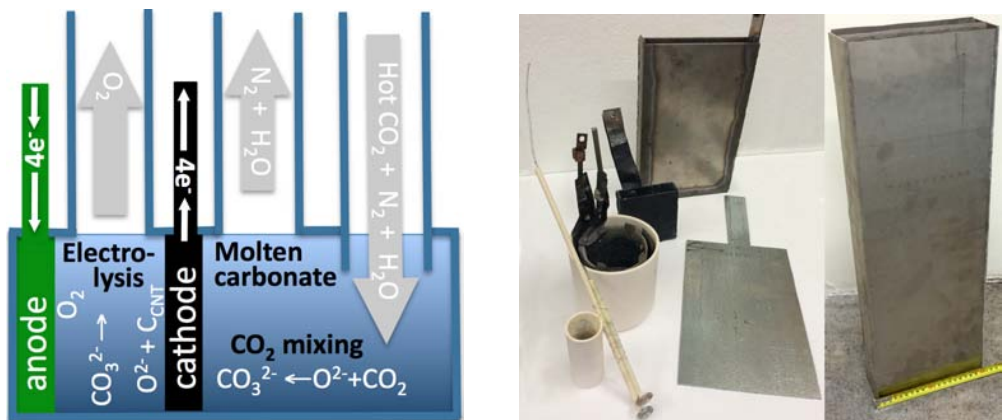

**Fig. 4.** Left: The $CO_2$ to CNT transformation, includes input of the $CO_2$ (and varying levels of $N_2$ and $H_2O$) gas, dissolving the $CO_2$ as molten carbonate, emitting any $N_2$ and/or $H_2O$ (which are insoluble in the molten carbonate), and splitting the $CO_2$ by electrolysis into carbon nanotubes at the cathode and releasing $O_2$ at the anode. Right: The evolution of the electrolysis chamber. Earlier versions can be seen in the front on the left, and later versions in the back and to the right. The rectangular electrolysis chambers use the interior walls of the nickel as air electrode.

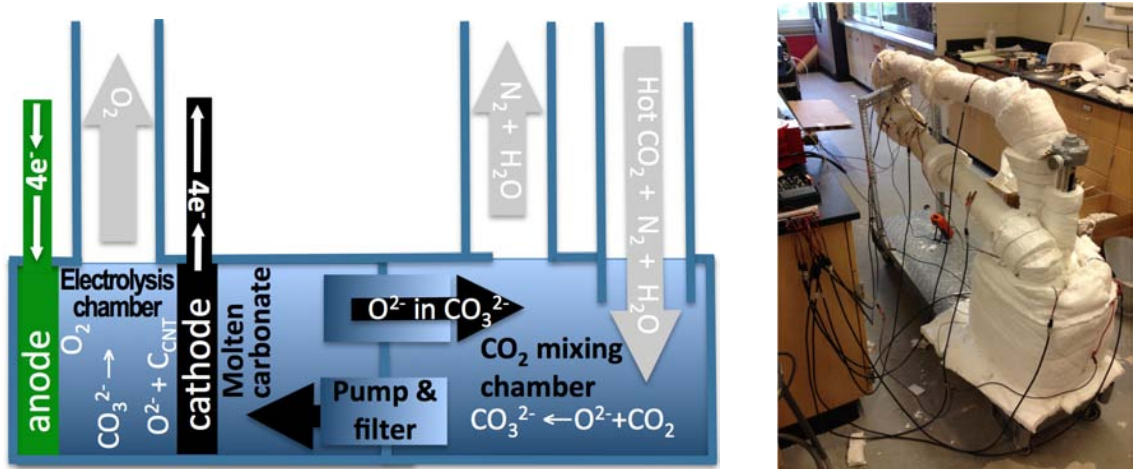

**Fig. 5.** Left: $CO_2$ to CNT transformation using a pumped, circulated molten carbonate electrolyte, a $CO_2$ mixing chamber inputting hot flue gas, dissolving the $CO_2$ and emitting the insoluble $N_2$ and $H_2O$ from a standard flue stack, and an electrolysis chamber producing carbon nanotubes at the cathode and releasing $O_2$ at the anode. Right: In-lab molten salt loop using a Wenesco HTPA2C molten salt pump.

**Cement production.** Society consumes over $3 \times 10^{12}$ kg of cement annually, and the cement industry releases 9 kg of $CO_2$ for each 10 kg of cement produced [1,2]. Cement $CO_2$ emissions are widely studied [18-22]. Components of a typical contemporary cement power plant are illustrated in Fig. 6. The plant feedstock is termed "meal" and typically consists of 75% limestone (calcium carbonate) and 25% clay (several silicates). Energy is conserved by blowing the exhaust gas back through the system in a series of preheaters (components 5 in the figure), and the stack emissions (components 1 & 10) can be combined in a single emission component. The rotary kiln partially melts the mix at 1450°C facilitating interaction between the oxide and silicates. At the lower end of the kiln (component 9), pulverized coal is used as the flame fuel, and fuel is also added to heat and drive the calciner and preheaters (component 6).

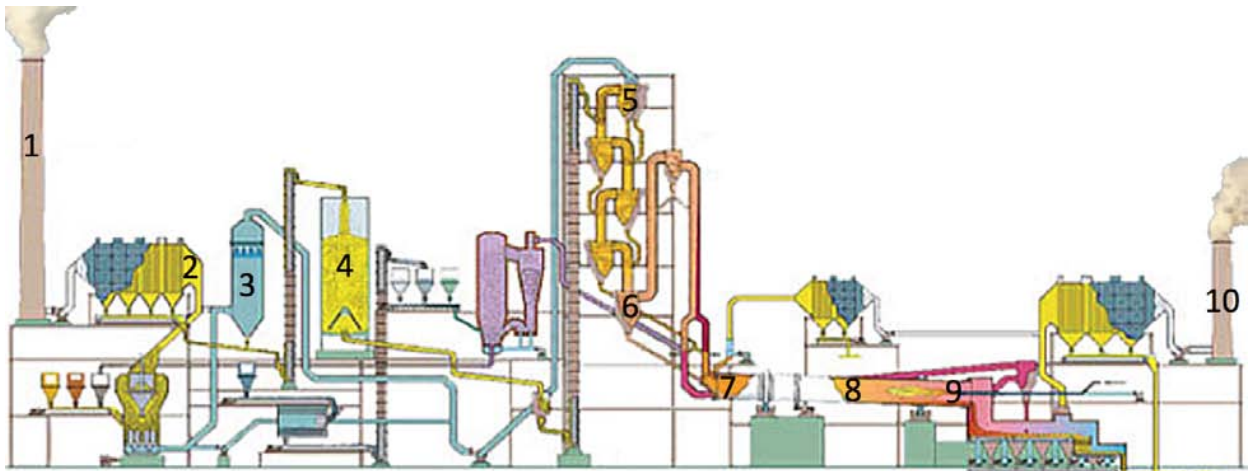

**Fig. 6.** Top: Components of a contemporary cement plant: 1) Main stack (Preheater & Calciner) with $CO_2$ emissions; 2) Electrostatic preciptator; 3) Flue gas scrubber; 4) Component mix; typically 75% limestone, 25% clay silicates; 5) Pre-heater; 6) Calciner; 7) Calcined mix rotary kiln inlet; 8) Rotary kiln; 9) Pulverized coal combustion; 10) Kiln stack with $CO_2$ emissions.
Adapted from http://www.e-inst.com/cement-furnaces-and-kilns/

An expanded illustration of a single preheater is shown in the middle of Fig. 7. A tower consisting of the preheaters and calciner is also included as a photo in the left side of the figure, and an example of the heating of the limestone mix that occurs as the kiln exhaust gas is blown back through the preheaters is illustrated on the right side of the figure.

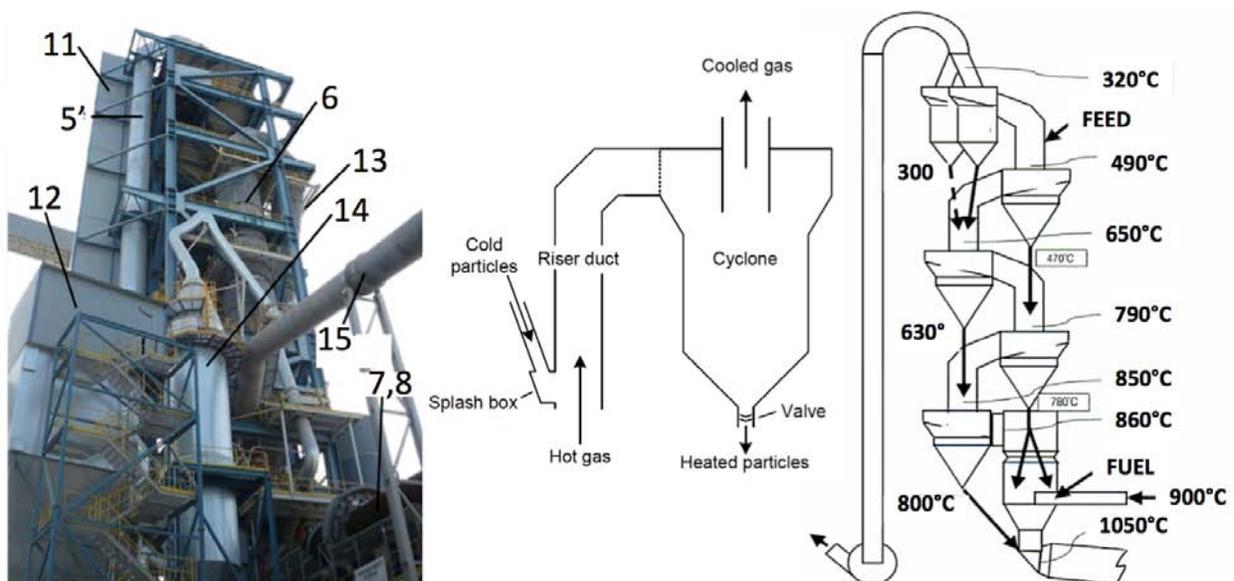

**Fig. 7.** An expanded view of the preheater tower in a cement plant. Left: photo of a 110 meter tall cement plant preheating and caliciner components of a cement plant including in addition to those labeled in Fig. 4: 5';) Gas duct leaving the preheater; 11) elevator and stairs; 12) Filter; 13) Downcomer; 16) Alkali bypass; 14 Tertiary duct. Total height of tower: 110 meters; middle preheating heat exchange; right: Four stage cyclone preheater with caliciner at bottom (typical temperatures shown).
Adapted from ref. 23.

In a contemporary cement power plant over half of the carbon dioxide produced is due to limestone to lime dissociation (eq. 2) and the remainder due to combustion of the fossil fuel, often coal (eq. 3), is to heat the feedstock:

$$CaCO_3 + Q_{heat} \rightarrow CaO + CO_2 \qquad (2)$$
$$nC + nO_2 \rightarrow nCO_2 + Q_{heat} \qquad (3)$$

A challenge to the economics of carbon capture at a cement plant is the relatively higher cost of capture compared to other industrial products. For example, the value per ton of cement produced is an order of magnitude less than the value per ton of steel, and hence the relative (as compared to product) price of carbon capture is substantially greater in the cement compared to the steel industry. A solution to this challenge is accomplished by the co-generation of a valuable product from the captured carbon, which can offset the carbon capture cost. The production of synthetic fuels from carbon dioxide (and/or water) has been widely studied and can be efficient. However, the value of such fuels including syngas and methane and hydrogen [15-17] is not significantly higher that of the cement product. In comparison, the highest grade industrial carbon nanotubes made from carbon dioxide are valued at ~1000 fold higher than such fuels, and provide substantial economic incentive to transform and eliminate the carbon dioxide emissions from cement plants.

Stack emissions from cement power plants contain ~30% carbon dioxide, which is a higher component than in fossil fuel power plant emissions. This higher concentration is due to the contribution of both the carbon dioxide from combustion and the carbon dioxide from the limestone to lime dissociation. Fig. 8 presents a block diagram of a conventional cement plant in which fuel (typically coal and/or natural gas [24]) and air are combined to generate the heat for the cement pyrolysis reactions, and the flue gas emissions contain a large volume of nitrogen from air, as well as carbon dioxide. Not shown in the diagram is if natural gas, principally methane, is used in heating, than the flue gas also contains steam:

$$nCH_4 + 2nO_2 \rightarrow nCO_2 + nH_2O + Q_{heat} \qquad (4)$$

As one pathway to decrease or isolate carbon dioxide emissions during cement production, oxy-fuel processes have been investigated [25-29]. In these processes, pure oxygen partially (Fig. 8B) or fully (Fig. 8C) replaces air for the fuel combustion, and hence the flue gas consists of a smaller volume then in the conventional cement production, and contains less nitrogen (Fig. 8B) or no nitrogen (Fig. 8C), and more carbon dioxide. In partial oxy-fuel processes the exhaust flue stream can contain up to 75% $CO_2$ and in full oxy-fuel processes over 90% $CO_2$. In partial oxy-fuel processes pure oxygen is typically considered for fuel combustion only during the calciner and preheating steps, but air is used during kiln combustion, whereas pure oxygen is used for both combustion sequences in full oxy-fuel processes. While the partial oxy-fuel process produces a lower concentration $CO_2$ stream, one advantage is that it is more applicable to retrofitting existing cement plants. The full ox-fuel process is more applicable to (redesigned) new power plants as the pure oxygen combustion in the kiln can lead to higher temperatures, which improve throughput rates, but can require alternative higher temperature resistant kiln materials.

Oxy-fuel processes can diminish greenhouse gas emissions through improved combustion efficiencies, but it should be noted that alone, oxy-fuel process concentrate, but do not capture, carbon dioxide and must be coupled with subsequent disposal of the concentrated $CO_2$ stream to achieve carbon capture. To date, a significant disadvantage of oxy-fuel processes has been the large energy and concurrent additional $CO_2$ emissions required to generate the pure oxygen. The principal technique to produce the requisite pure oxygen has been via cryogenic cooling and separation from the liquefied air, but also solid membranes to separate $O_2$ from air have been explored [29]. Both of these methodologies are energy intensive, which increases the cost of $CO_2$ extraction, and diminishes the extent to which net $CO_2$ can be isolated during the cement production. However, both partial oxy-fuel combustion for retrofitting existing cement plants, and full oxy-fuel combustion for building new cement plants, are considered to be amongst the most promising technologies for carbon sequestration during cement production.

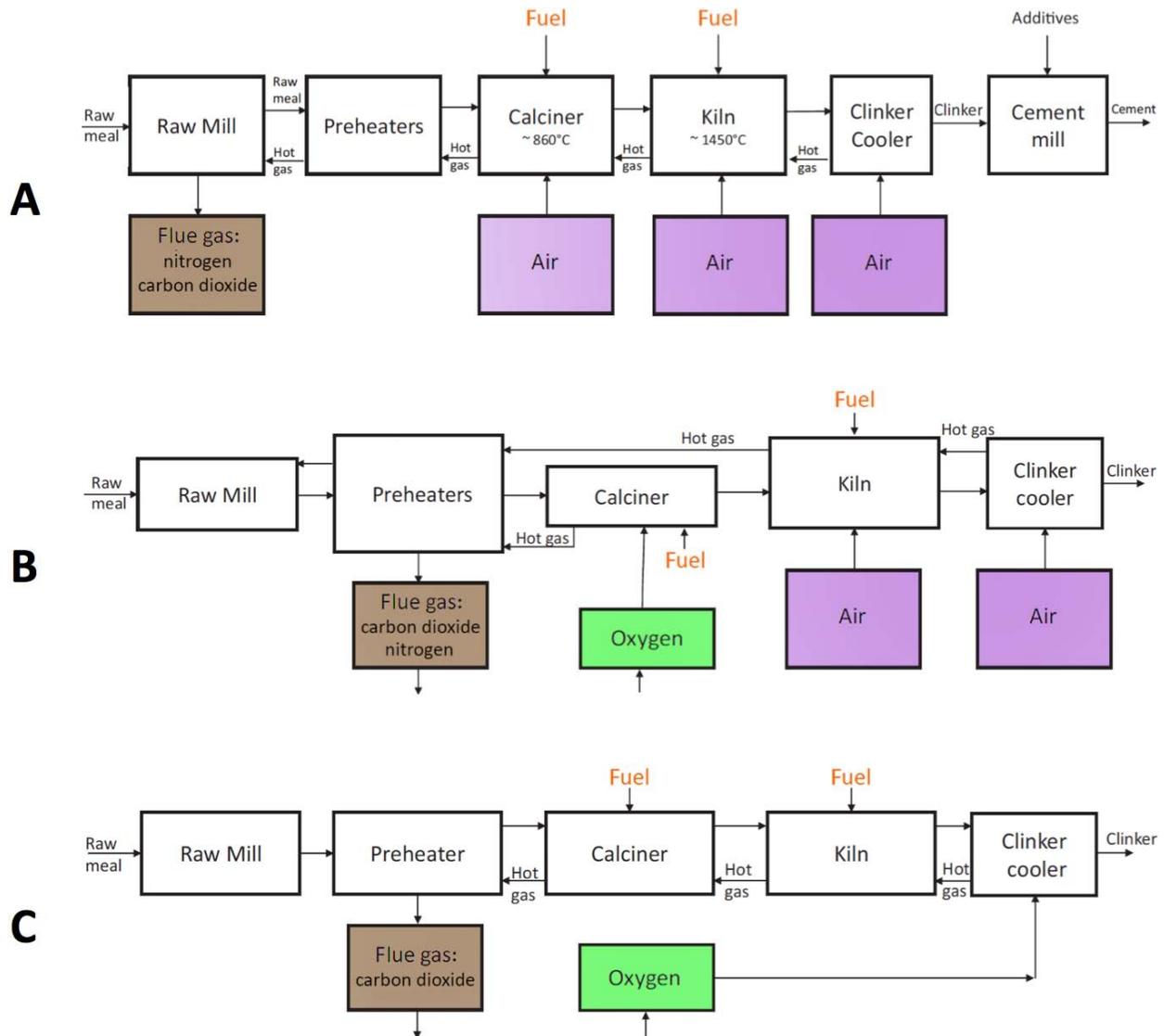

**Fig. 8.** Comparison of alternative cement production process. A: Conventional configuration – block diagram; B: Partial oxy-fuel configuration; C: Full oxy-fuel configuration. Adapted from reference 25.

**Coupling Cement and C2CNT processes.** $CO_2$ electrolytic transformation to carbon nanotubes for cement plants as introduced in this study has several advantages for elimination of this greenhouse gas. (i) Rather than just formed into a concentrated stream, the carbon dioxide is captured (as carbon nanotubes), (ii) pure oxygen is produced as a byproduct of the carbon dioxide splitting and available for oxy-fuel production without an additional energy cost to improve the plant energy efficiency and production rate, and (iii) the carbon nanotubes are valued at over 1000 fold that of cement providing an economic incentive for removal and transformation of the $CO_2$ emission stream during cement production.

**The C2CNT cement process.** Fig. 9 illustrates coupling of cement production to the C2CNT process. Rather than producing a $CO_2$ emission, the plant gas enters the C2CNT electrolysis chamber, is converted to carbon nanotubes and the flue gas emission contains no anthropogenic carbon dioxide. In both the partial (figure left) or full (figure right) oxy-fuel configurations by removing air, the flue gas has a smaller volume than in conventional cement plants. Oxygen generated in the C2CNT electrolysis chamber loops back into the cement line improving the combustion efficiency (heat delivered) of the fuel and decreasing plant gas volume decreasing radiative heating and increasing the rate of meal processing.

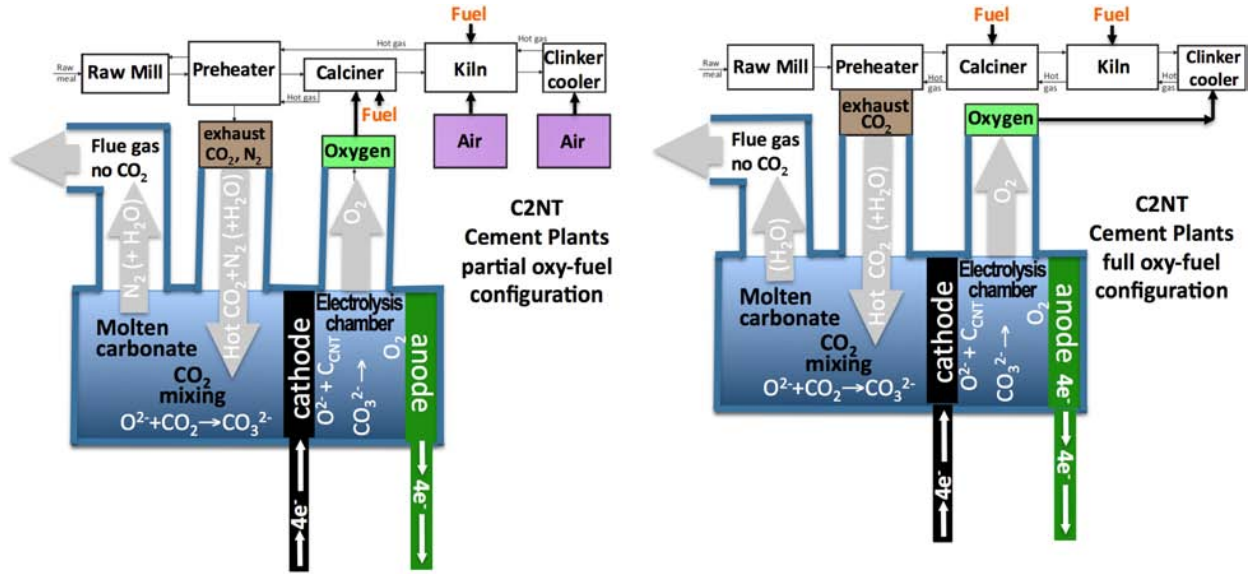

**Fig. 9.** C2CNT Cement: The coupling of C2CNT to a cement plant.

In the full oxy-fuel configuration illustrated on the right side of Fig. 9, the oxygen originating from the electrolytically split carbon dioxide is looped back through the clinker cooler and kiln, and external air is not required during meal conversion to cement. The flue only has a significant volume if methane (rather than pure coal) is included during fuel combustion as this generates, in which case in addition to carbon dioxide, one water per methane combusted (eq. 4) is produced, and this water vapor is vented as flue gas. Such water vapor is insoluble in the C2CNT molten carbonate and exits as flue gas. In the partial oxy-fuel configuration illustrated on the left side, the C2CNT oxygen is looped back through the calciner, but air is added and combusted in the kiln. This partial addition of air releases nitrogen (but a smaller volume than in conventional cement plants) during kiln combustion, which is subsequently emitted (as a smaller volume) flue gas. While the full oxy-fuel C2CNT cement configuration offers energy savings and process acceleration advantages due to the lower volume of gas processed. The partial oxy-fuel process is generally considered easier to adapt to existing cement plants through a small retrofit of the calciner, while kiln modifications, and higher heat resistant materials to optimize the full oxy-fuel C2CNT cement process may be applicable to new plants.

The C2CNT cement configurations illustrated in Fig. 9 are $CO_2$-free when a non-$CO_2$ emitting electrical source (nuclear or renewable) is used to drive the C2CNT electrolysis. The global C2CNT cement process is carbon negative as the cement produced absorbs atmospheric $CO_2$ over time (whose rate depends on the cement mixture and curing conditions), generalized and simplified as the spontaneous process:

$CaO + CO_2 \rightarrow CaCO_3 + Q_{heat}$ (5)

Renewable energy electricity prices are dropping as the various wind, solar and hydroelectric power plant technologies continue to mature. The least expensive renewable energy today appears to be wind power generated in Texas, USA, at a cost of $0.029 kWh$^{-1}$ (and even lower annual average costs of $0.025 have been quoted online in references 30 and 31. At high coulombic yields (4e- per $CO_2$), a 1 V electrolysis costs $70 per ton $CO_2$ processed (determined as $0.029 kAh$^{-1}$ x 22.7x10$^3$ mol ton$^{-1}$ x 4 e- x F, F=Faraday's constant). However the cost is less as it also includes the co-generation of oxygen in the C2NT chamber. The actualized electrical cost of $50 per ton $CO_2$ is estimated by inclusion of the concurrent large oxy-fuel energy efficiency improvement of the cement plant, compared to conventional air-driven configuration. C2CNT capital costs are also low. The C2NT electrolysis chamber does not use noble metals. The nickel anode, steel cathode and carbonate electrolyte is not consumed and the costs are low when amortized over the plant life. In addition to pure lithium carbonates, we have demonstrated that less expensive sodium and calcium carbonate mix electrolytes also facilitate $CO_2$ to carbon nanotube electrolysis at high yield. These costs compare to the value of the C2CNT cement products. The 1.1 ton cement per ton $CO_2$ produced is typically valued at ~$110, and the process also generates industrial grade pure carbon nanotubes, CNTs. In accord with the formula weights, each 44 kg of $CO_2$ generates 12 g of CNT, or 0.27 ton of CNT per ton of $CO_2$. Industrial grade CNTs are currently valued at ~$225,000 ton$^{-1}$ (varying with purity and properties). Hence a C2CNT cement plant emits no $CO_2$, consumes ~$50 of electricity and produces ~$100 of cement and $60,000 of carbon nanotubes per ton of $CO_2$ avoided. A net profit in excess of over $50,000 per ton of $CO_2$ avoided provides a powerful incentive to mitigate this greenhouse gas. The market for C2CNTs includes the $trillion dollar market as lightweight replacement for the existing iron and aluminum market, and other markets include packaging, transportation, building materials, batteries and nanoelectronics. The value per ton of CNTs will fall in time, which is expected to accelerate market growth. Even if an order of magnitude drop or more in CNT value occurs with market growth, the marginal profit remains substantial and the C2CNT cement plant provides an economic incentive, rather than economic cost, to mitigate climate change through a carbon negative process.

We have previously demonstrated wind turbine electric powering molten carbonate electrolyses [32], and here a wind powered C2CNT cement configuration is illustrated in Fig. 10.

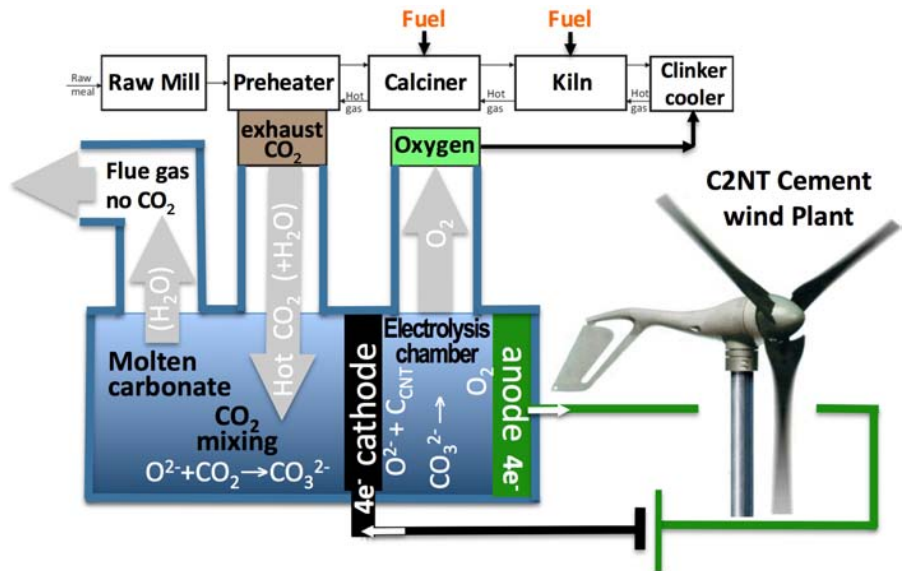

**Fig. 10.** C2CNT Cement wind plant: The full oxy-fuel configuration is shown. The plant is does not emit $CO_2$, and over time cement produced absorbs $CO_2$. Hence the process is carbon negative, which compares favorably to the large positive carbon signature of conventional cement plants.

**Heat management in the C2CNT cement plant.** The heat in the C2CNT electrolysis chamber is a balance between the temperature of the cement plant exhaust feedstock into the C2CNT chamber, $T_{CO2\text{-}gas}$, the heat of dissolution of the $CO_2$ in the molten carbonate electrolyte, $Q_{dissolution}$, radiative and heat exchange losses, $Q_{radiative}$, and the heat of electrolysis while splitting $CO_2$ at an applied electrolysis potential, $Q_{overpotential}$. $T_{CO2\text{-}gas}$ will depend on the point at which the $CO_2$ exhaust feedstock is extracted from the cement line. A likely extraction point is the ~300°C exhaust gas from the upper most preheater on the right side of Fig. 7. An alternative extraction point is the ~150°C at the point of clinker cooling. It will be shown that the additional 450 to 600°C of heating (for an electrolysis sustained at 750°C) can be supplied by overpotential and dissolution heating during the C2CNT electrolysis, and that there is an opportunity for excess heat generation by electrolysis to improve the fuel efficiency of the plant calcination process.

Variation of the redox potential (from the free energy) and the thermoneutral potential (from the enthalpy) of carbon dioxide splitting by electrolysis to either a carbon monoxide or carbon (plus oxygen) product is presented in Fig. 11. The potentials are calculated from the thermochemical from the enthalpies and enthalpies of the individual $CO_2$, C, CO and $O_2$ species []. As can be seen in the figure, the thermoneutral potential for both the CO and C products is approximately constant over a wide temperature range. Whereas the redox potential for the carbon monoxide product is exogenic, decreasing rapidly with increasing temperature, the redox potential for the carbon product is highly constant despite changing temperature.

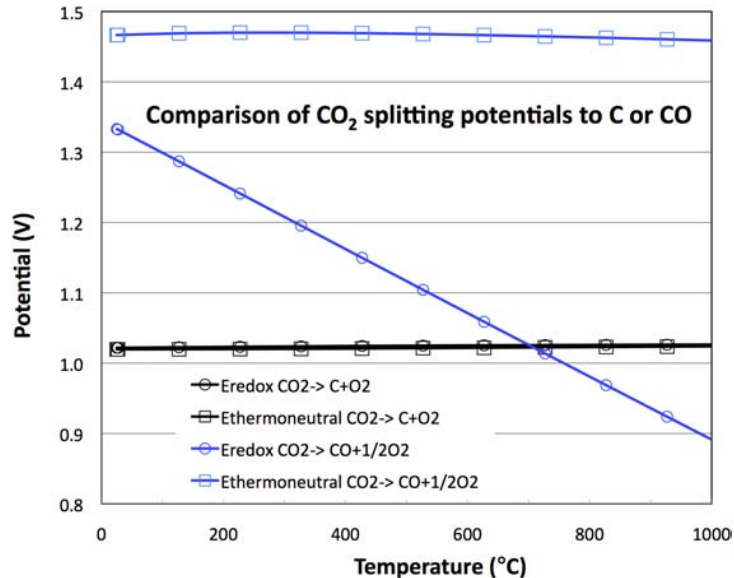

**Fig. 11.** The redox and thermoneutral potentials) of carbon dioxide splitting by electrolysis to either a carbon monoxide or carbon (plus oxygen) product. Calculated using themochemical values of the individual species [33, 34].

As seen in Fig.11, below 800°C the thermodynamic potentials for carbon dioxide splitting to carbon has a same consistent thermodynamic redox and thermoneutral potential of 1.02 V. Above 800°C, in accord with the lower thermodynamic CO potentials at higher temperature in the figure, we observe with increase in temperature that the product smoothly shifts from pure carbon at 750°C to pure CO at 950°C, We have demonstrated that the electrolysis potential for splitting of carbon dioxide in molten carbonates further decreases with increasing concentrations of added oxide, and full cell potentials of < 0.75 V are measured for carbon grown by electrolysis in 4 m $Li_2O$ in $Li_2CO_3$ at 750°C [7].

Molten carbonates used as the electrolyte in C2CNT electrolysis exhibit melting points as low as 399°C (for the lithium, sodium potassium carbonate eutectic), 723°C for pure Li2CO3, and the melting point is below 700°C for Li/Na and Li/Ba carbonate eutectic mixes. Increasing applied potentials above the redox potential generate increasing electrolysis current densities. Applied potentials above the thermoneutral potential generate heating of the system (and applied potentials below the thermoneutral potential lead to system cooling). In a 4 e- process (n=4), the heat generated by an 0.1, 0.2 or 0.5 V overpotential (from $\Delta E_{overpotential} = Q_{overpotential}/nF$; $Q_{overpotential} = \Delta E_{overpotential} *4*96.485$ kJ $V^{-1}$ $mol^{-1}$) is respectively: = 38.6, 99.2 and 193.0 kJ $mol^{-1}$ $CO_2$. We observe that for a typical molten carbonate such as $Li_2CO_3$, s = 2 mole of electrolyte is sufficient to dissolve 1 mole of $CO_2$. Molten $Li_2CO_3$ has a highly invariant heat capacity of $C_p$ = 0.1854 kJ / mol °C. Hence, $Q_{overpotential}$ is sufficient to provide an additional heating of the system, beyond that of the temperature of the $CO_2$ feedstock, of $\Delta T = Q/sC_p$ = 104°C, 268 or 520°C respectively at electrolysis overpotentials of 0.1, 0.2 or 0.5 V.

Furthermore in accord with eq 1, the chemical dissolution of $CO_2$ by $Li_2O$ in $Li_2CO_3$ releases $Q_{dissolution}$ = 158 kJ $mol^{-1}$ of heat. If the combined heats of $Q_{overpotential}$ and $Q_{dissolution}$ are in excess of radiative losses than C2CNT will add heat that will decrease the heat by fuel required for calcination, and further improve the energy efficiency of the cement plant.

**Advanced C2CNT cement calicination-free plants**. The C2NT process described in Figs. 4 and 5 is an indirect process as described on the left side of Fig. 12, in that $CO_2$ from $CaCO_3$ calicination is electrolysed to form carbon. As one alternative, we have also demonstrated the direct electrolysis of $CaCO_3$, which is highly soluble in $Li_2CO_3$ as measured in Fig. 13 on the left. This alternative process, which eliminates the calcination step is shown on the right side of the figure. As in the indirect C2CNT process, the electrolysis produces oxygen at the anode, and simultaneously produces not only carbon, but also lime (CaO) for cement at the cathode:

Dissolution & Electrolysis:   $CaCO_3$(soluble) → CaO(insoluble) + C(insoluble) + $O_2$(gas)     (6)

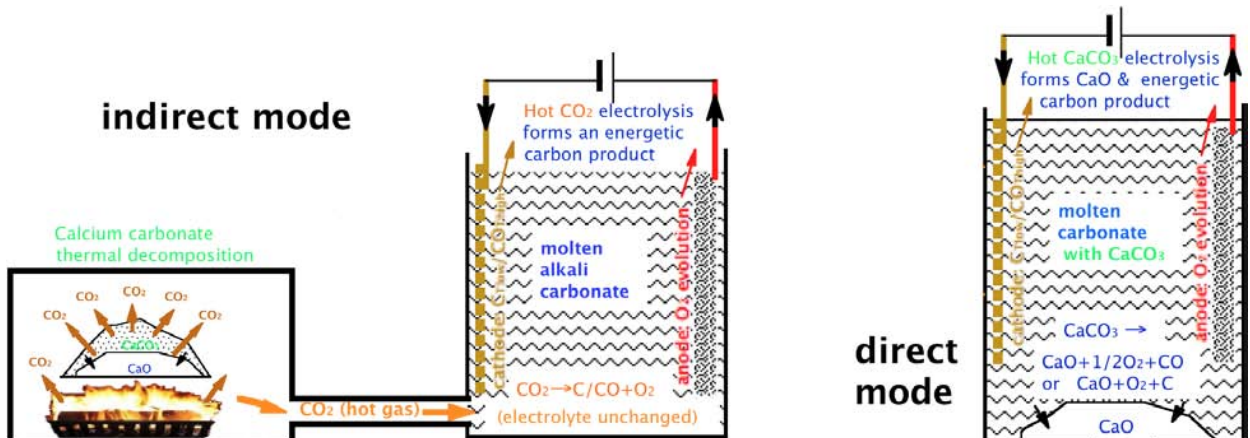

**Fig. 12.** Indirect versus direct electrolytic production of lime from limestone. Left: The C2CNT cement process delineated in prior sections of this study is indirect process in which CaO is formed in one area by thermal calcination, and the exhaust is fed into a separate electrolysis chamber in which the $CO_2$ is transformed into oxygen and carbon.  Right: In the alternative direct mode no thermal calcination is required and $CaCO_3$ (dissolved in molten carbonate) is directly converted by electrolysis into oxygen, carbon and CaO. Adapted from ref. [13].

A second alternative (not shown in Fig. 12) also avoids thermal calcination and operates in single electrolysis chamber, However, neither $CO_2$ nor $CaCO_3$ is added to the molten carbonate during the electrolysis, This electrolysis splits another molten carbonate, such as $Li_2CO_3$ to $O_2$, C and soluble $Li_2O$. Subsequently soluble $CaCO_3$ is added which spontaneous reacts with the soluble $Li_2O$ forming insoluble CaO which precipitates for product extraction:

Electrolysis: $Li_2CO_3$(molten) → C(insoluble) + $Li_2O$ (soluble) + $O_2$(gas)
Dissolution & Precipitation: $Li_2O$(soluble)+$CaCO_3$(soluble) →CaO(insoluble)+$Li_2CO_3$(molten)
Net: $CaCO_3$(soluble) → CaO(insoluble) + C(insoluble) + $O_2$(gas)     (7)

Unlike, lithium oxide, calcium oxide is highly insoluble in the molten carbonate electrolyte as measured on the left side of Fig. 13. The insoluble CaO, produced in a 1:1 molar ratio with carbon during the electrolysis, is more dense than the electrolyte, and under various conditions this lime tends collect at the cathode or to precipitate at the bottom of the electrolysis chamber. Measured limestone splitting potentials are presented in Fig. 13 on the right and it should be noted that the high current densities (in excess of 1 A cm$^{-2}$) provide high limestone throughput and rapid conversion of limestone to lime and carbon.

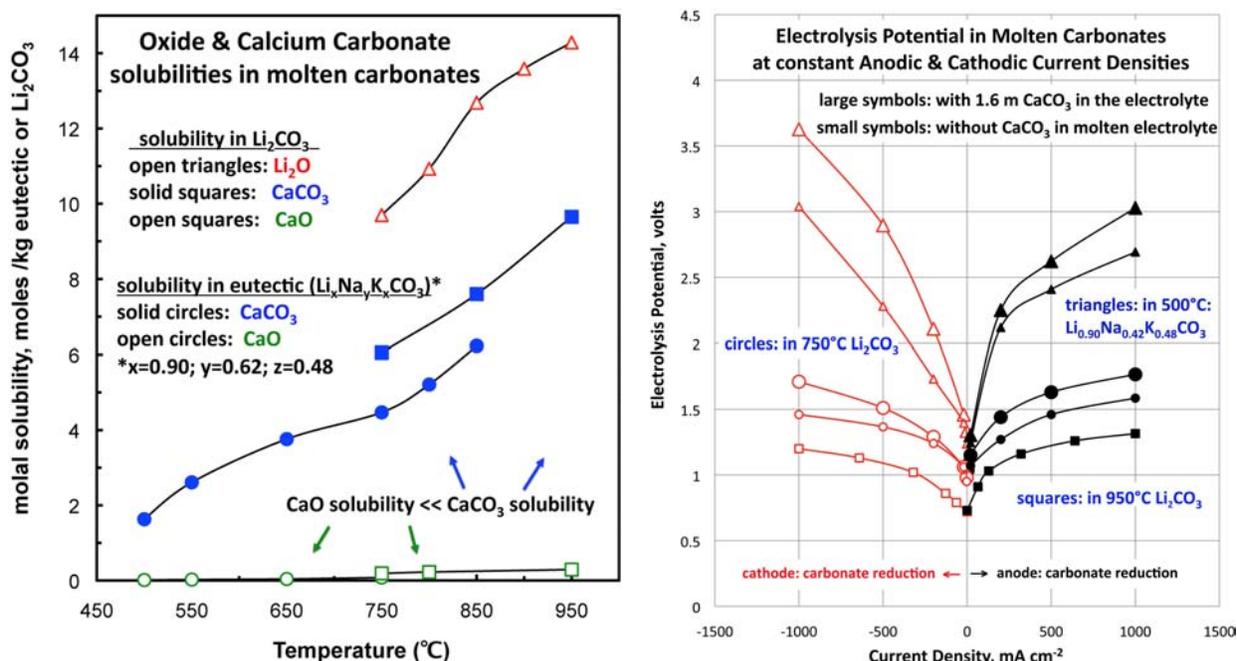

**Fig. 13.** Top: The low solubility of calcium oxide, compared to calcium carbonate and lithium oxide solubility in molten carbonates (top) facilitates the electrolysis and precipitation of calcium oxide. Bottom: The measured full electrolysis potential as a function of current density in either $Li_2CO_3$ at 750 or 950 °C, or eutectic molten carbonates at 500 °C at an iron cathode or nickel anode (with excess oversized counter electrodes). From ref. [13].

X-ray powder diffraction (Fig. 14 top) and FTIR (Fig. 14 bottom) analysis of the CaO product indicates the CaO is of high purity [13]. We have recently demonstrated that carbon nanotubes are produced from this molten $CaCO_3$/$Li_2CO_3$ mixed electrolyte, such as by electrolysis on a steel cathode with 20 wt% $CaCO_3$ added to a molten $Li_2CO_3$ electrolyte [10].

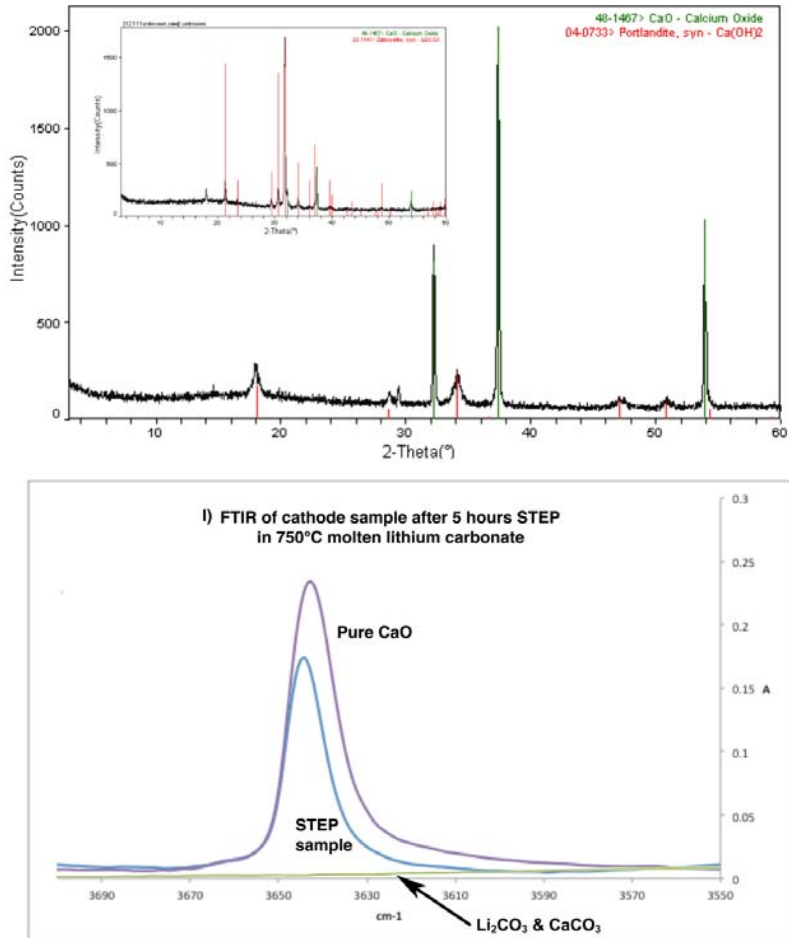

**Fig. 14**. Analysis of CaO products from $CaCO_3$ following a 5 hour electrolysis in 750°C $Li_2CO_3$, Left: XRD of CaO during bulk electrolytsis of carbonate. Inset: CaO product removed nearer to the electrolyte interface additionally carries clean electrolyte ($Li_2CO_3$). Library reference spectra for CaO, $Ca(OH)_2$ and $Li_2CO_3$ are indicated as vertical lines. Right: CaO FTIR absorption spectrum of cathode deposited CaCO product following the electrolysis, and following base line correction to remove broad band carbon absorption. From ref. [13].

**Conclusion**. C2CNT (Carbon dioxide to carbon nanotube) cement plants have been introduced and analyzed which provide a significant economic incentive to eliminate the massive $CO_2$ greenhouse gas emissions of current plants and serves as a template for carbon mitigation in other industrial manufacturing processes. Rather than regarding $CO_2$ as a costly pollutant, this is accomplished by treating $CO_2$ as a feedstock resource to generate valuable products (carbon nanotubes). The exhaust from partial and full oxy-fuel cement plant configurations are coupled to the inlet of a C2CNT chamber in which CO2 is transformed by electrolysis in a molten carbonate electrolyte at a steel cathode and an nickel anode. In this high yield 4e- per $CO_2$ process, the $CO_2$ is transformed into carbon nanotubes at the cathode, and pure oxygen at the anode that is looped back in improving the cement line energy efficiency and rate of production. A partial oxy-fuel process looping the pure oxygen back in through the plant calcinator has been compared to a full oxy-fuel process in which the pure oxygen is looped back in through the plant kiln. The partial oxy-fuel process provides the advantage of easier retrofit of existing power plants, while the full oxy-fuel process provides the advantage of maximum efficiency by minimizing the volume of gas throughput (eliminating nitrogen from air).

An upper limit to the electrical cost to drive C2CNT electrolysis is $70 based on Texas wind power costs, but will be lower with fuel expenses when oxy-fuel plant energy improvements are taken account. The current value of a ton of carbon nanotubes is substantially in excess of a ton of cement. Hence a C2CNT cement plant consumes ~$50 of electricity, emits no $CO_2$, and produces ~$100 of cement and $60,000 of carbon nanotubes per ton of $CO_2$ avoided. A net profit in excess of over $50,000 per ton

of $CO_2$ avoided provides an incentive to mitigate this greenhouse gas. Even with a significant drop in CNT value with market growth, the C2CNT cement plant provides high marginal cost profits with $CO_2$ elimination from the plant. This is a powerful economic incentive, rather than economic cost, to mitigate climate change through a carbon negative process.